\documentclass[%
 reprint,
superscriptaddress,
 amsmath,amssymb,
 aps,
]{revtex4-2}

\usepackage{graphicx}
\usepackage{dcolumn}
\usepackage{color}
\usepackage{bm}

\makeatletter
\def\@author@present@script#1#2#3{%
  \begingroup
    \gdef\comma@space{\textsuperscript{}}
    \doauthor{#2}{#3}{\@affil@present@script}%
  \endgroup
  \advance\@tempcnta\m@ne
}
\makeatother

\begin{document}

\title{{Directed swimming of \textit{Chlamydomonas reinhardtii} near complex microstructures}}

\author{Chunhe Li}
\affiliation{Global College, Shanghai Jiao Tong University, Shanghai 200240, People's Republic of China}
\author{Hongyi Bian}
\affiliation{Global College, Shanghai Jiao Tong University, Shanghai 200240, People's Republic of China}
\author{Zixiang Lin}
\affiliation{Global College, Shanghai Jiao Tong University, Shanghai 200240, People's Republic of China}
\author{Yi Man}
\affiliation{Department of Mechanics and Engineering Science, College of Engineering, 
Peking University, Beijing 100871, People's Republic of China}
\author{Zijie Qu}
\thanks{Contact author: \href{malto:zijie.qu@sjtu.edu.cn}{zijie.qu@sjtu.edu.cn}}
\affiliation{Global College, Shanghai Jiao Tong University, Shanghai 200240, People's Republic of China}

\begin{abstract}
The locomotion of microorganisms near solid-liquid interfaces is of significant scientific interest due to its relevance in various natural and industrial contexts, including biofilm formation and marine biofouling. In this study, we investigate the swimming behavior of $C.\ \textit{reinhardtii}$ near a sinusoidal wave-like microstructure. Using fluorescence microscopy and a three-dimensional tracking technique, we observe that the swimming direction of $C.\ \textit{reinhardtii}$ is strongly influenced by the geometric constraints imposed by the microstructure. Our results reveal that cells preferentially accumulate at the valley of the microstructure rather than the peak and exhibit a directional swimming tendency in the vicinity of the microstructure. We attribute this behavior to a combination of the “memory effect” and hydrodynamic attraction. By modifying the geometry of the microstructure, we successfully achieve directed guidance of cell swimming, which demonstrates potential applications in micro and nanorobotics and antifouling technologies. This work provides alternative insights into the locomotion mechanisms of microorganisms near solid-liquid interfaces and underscores the potential for manipulating their behavior through microstructure design.

\end{abstract}

\maketitle


\section{\label{sec:level1}Introduction}

Microorganism locomotion near interfaces plays a crucial role in a wide range of natural and engineered processes, from the colonization of medical implants and marine biofouling to fertilization and microbial transport in porous media~\cite{KRISHNAMURTHI2022112190,https://doi.org/10.1002/admi.202201425,PhysRevE.109.054601,PhysRevE.99.052607}. Although the hydrodynamic, steric, and flagellar interactions between microorganisms and flat solid–liquid interfaces are now relatively well understood, leading to behaviors such as accumulation, hydrodynamic trapping, and reorientation, real-world boundaries are far more complex~\cite{C3SM51426A,scheidweiler2024spatial,malkin1996bill}. The micro and nanoscale features will dramatically alter local flow fields, introduce geometric constraints, and generate novel guidance or trapping effects~\cite{10.1104/pp.42.9.1284,PhysRevLett.128.258101}. Therefore, understanding how microorganisms navigate these complex surfaces is crucial for elucidating microbial behavior in realistic environments and designing engineered structures that can control, harness, or inhibit their motion. 

Existing studies have reported phenomena such as directional rectification and preferential accumulation~\cite{PhysRevFluids.5.073103,kurzthaler_microswimmers_2021}. Using a small-surface-roughness perturbation approach, Kurzthaler \textit{et al.} demonstrated that anisotropic pressure gradients induced by surface corrugations drive transverse flows. Furthermore, by tracking the motion of tracer particles, three-dimensional helical trajectories were observed, revealing an overall drift along the direction of the corrugations~\cite{Kurzthaler_Chase_Stone_2024}. Ishimoto \textit{et al.} investigated the motility of swimmers over single- and double-periodic sinusoidal surface topographies through numerical simulations. When the swimmer size is comparable to the surface wavelength, swimmers can be guided to move along the troughs of the topography; however, this guiding effect disappears when the wavelength is much larger or much smaller than the swimmer size~\cite{ishimoto2023squirmer}. Beyond fundamental understanding, these findings have paved the way for various applications, including bioinspired microrobotics~\cite{ceylan_3d-printed_2019,doi:10.1126/scirobotics.aar4423,https://doi.org/10.1002/adhm.202401383}, microfluidic device designing~\cite{10.7554/eLife.85348,doi:10.1073/pnas.2013925118,annurev:/content/journals/10.1146/annurev.fluid.36.050802.122124}, and antifouling surface engineering~\cite{CHOI2023121814}. 

\begin{figure*}
\centering
\includegraphics[width=1.0\linewidth]{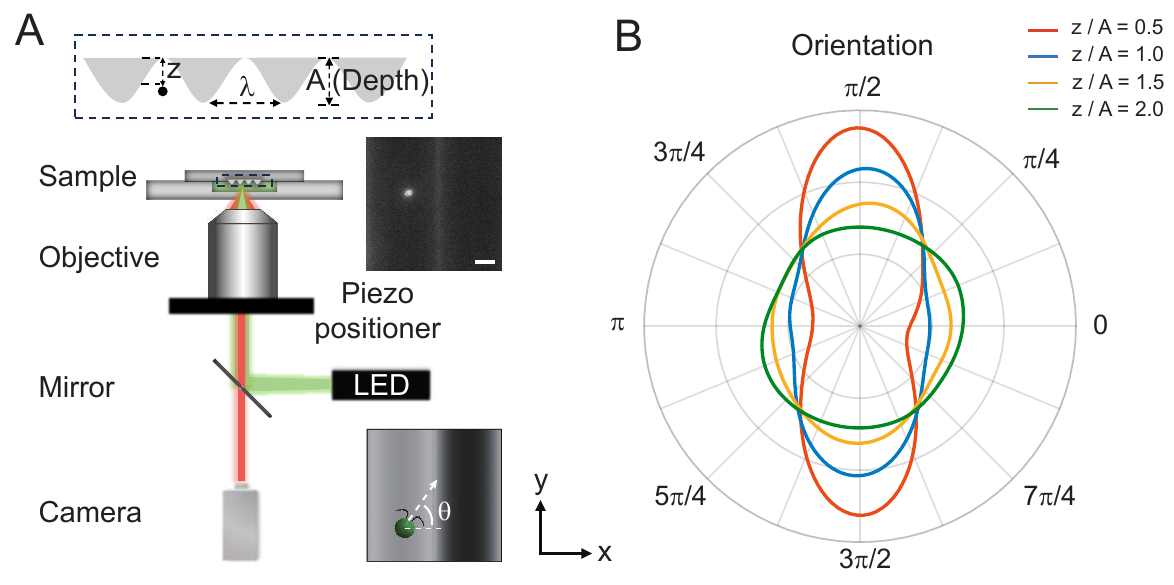}
\caption{\label{fig1}(A) {Schematic diagram of the experimental setup. The suspension of $C.\ \textit{reinhardtii}$ is injected into the chamber, above which is a glass slide with a patterned surface of wavelength $\lambda = 50~\mu\mathrm{m}$ and depth $A = 50~\mu\mathrm{m}$.} The cells are illuminated with a light source of 565 nm wavelength. The filter retains the fluorescence emitted by the cells near 610 nm. {The piezoelectric positioner is operated in a closed-loop feedback mode to maintain the cell near the focal plane, enabling real-time determination of the three-dimensional motion trajectory.}~More details are shown in Supplemental
Material~\cite{Supplemental Material}. Inset schematically show fluorescent image of one cell swimming near the microstructure and schematic diagram of the angle between the swimming direction of cell and x-axis. (B) Under different distances towards the bottom of the microstructures, the angle distribution between the cell's motion direction and the positive direction of the x - axis. z is the distance between the cell and the top substrate, and A represents the depth of the microstructure. As the distance gradually increases, the cell swimming orientation distribution changes from being concentrated at $\pi/2$ and $3\pi/2$ to a uniform distribution. Scale bar: 20 $\mu m$.}
\end{figure*}

Among the diverse microbial systems studied, the biflagellated green algae \textit{Chlamydomonas reinhardtii} has emerged as a widely adopted model for investigating swimming motility at interfaces~\cite{zeng_sharp_2022,yuan_hydrodynamic_2015,PhysRevE.82.056309,kantsler_ciliary_2013}. Its well-characterized flagellar mechanics, robust phototactic behavior, and intrinsic autofluorescence make it particularly suitable for quantitative studies of microscale motility~\cite{polin_chlamydomonas_2009,jeanneret_brief_2016,PhysRevLett.109.138102}. Previous research has shown that \textit{C.\ reinhardtii} exhibits surface scattering dynamics and hydrodynamic interactions that lead to accumulation phenomenon~\cite{lushi_scattering_2017}. Ostapenko \textit{et al.} examined the motility of \textit{C.\ reinhardtii} in microhabitats with controllable geometric features. They found that the geometry of the habitat governs cellular navigation in confined environments: the probability of cells swimming near boundaries increases with wall curvature~\cite{ostapenko_curvature-guided_2018}. Contino \textit{et al.} conducted the first detailed investigation of the scattering behavior of \textit{C.\ reinhardtii} from convex obstacles, revealing the interplay between hydrodynamic and contact forces at different incidence angles. They proposed a classification framework distinguishing stochastic from deterministic scattering and quantified the corresponding probability distributions~\cite{contino_microalgae_2015}. These characteristics, together with its well-established experimental tractability, make \textit{C.\ reinhardtii} an ideal model organism for elucidating the fundamental physics of microswimmer–boundary interactions. 

Prior studies in quasi-2D chambers elegantly linked curvature to near-wall probability, typically relying on steric/Brownian frameworks without hydrodynamic interactions. Yet quasi-2D geometries restrict out-of-plane reorientation and limit access to 3D guidance phenomena. In this study, we close this gap~\cite{ostapenko_curvature-guided_2018,PhysRevE.99.052607}. We use fluorescence 3D tracking techniques~\cite{qu_changes_2018} to systematically examine the swimming behavior of \textit{C.\ reinhardtii} near sinusoidal microstructures. An obvious alignment of cell swimming trajectories along the grooves of the microstructure is observed, with the orientational bias extending beyond half of the microstructure depth. Furthermore, cells prefer to accumulate in the troughs rather than crests of the microstructure, highlighting a spatial bias in residence time distribution. To elucidate the underlying mechanisms, we quantify the magnitude of components of swimming velocity and analyze the swimming orientation distribution of trajectories at the entrance and exit of the microstructure. The results demonstrate that the orientation of cells near the microstructure is influenced not only by random motion but also by hydrodynamic interactions. For the sake of simplicity, we approximate the sinusoidal profile as a square-wave-like geometry. This allows us to apply our previously developed pusher-puller hybrid model for further analysis~\cite{PhysRevE.111.L052401,buchner_hopping_2021}. This, combined with simulations of wall-induced scattering, shows that both direct collisions and hydrodynamic flows play synergistic roles in generating the observed orientation bias. Based on these results, we design microstructures of a specific shape to achieve structure-guided directional swimming of \textit{C.\ reinhardtii}. Our work sheds light on control strategies for directed swimming of microorganisms and provides new insights for the design of microfluidic lab-on-chip devices.

\section{\label{sec:level1}Result and Discussion}
\subsection{\label{sec:level2}Swimming orientation distribution near the microstructure}
To investigate the swimming behavior of cells near the surface with complex microstructure, we design and manufacture an observation chamber as shown in Fig. \ref{fig1}A. A glass slide with a sinusoidal periodic microstructure is placed on top of the observation chamber with a height of 500 $\mu m$ from the peak of microstructure to the lower surface of chamber, and the other two dimensions of the chamber are approximately 15 mm, so the side wall effects is negligible. The surrounding area is sealed with vaseline to ensure that the chamber is completely filled with the cell suspension. Cells are imaged using an inverted fluorescent microscope. (See Materials and Methods Fluorescence Tracking Microscopy).

By recording two-dimensional swimming trajectories of cells at different focal planes, we measure the angle between the cell swimming direction and the positive \textit{x}-axis, as shown in Fig. \ref{fig1}B. The microstructure expanded along the x direction with no structure along the y direction. It is found that the cell swimming orientation is closely related to the distance from the microstructure. The dimensionless parameter $z/A$ is used to represent the position of cells near the microstructure, where $z$ denotes the distance from the cell to the valley of the microstructure (as shown in Fig.~\ref{fig1}A), and $A$ is the depth of the microstructure. It is observed that for $z/A = 0.5$ and $1.0$, where the cells are located inside the microstructure, the angles between the cell swimming direction and the positive x-axis are predominantly concentrated around $\pi/2$ and $3\pi/2$, indicating a tendency for the cells to swim along the $y$-axis. This result suggests that the swimming behavior of the cells may be influenced by the hydrodynamic interaction and the bounded effect imposed by the microstructure. Besides, it is worth noting that for $z/A$ = 1.5 (i.e. cells are located outside the microstructure), the direction bias of the swimming cell still exists. Until $z/A$ = 2.0, the swimming orientation distribution tends to be uniform.

\subsection{\label{sec:level2}Residence time of cells at different positions near the microstructure}
\begin{figure}
\centering
\includegraphics[width=1.0\linewidth]{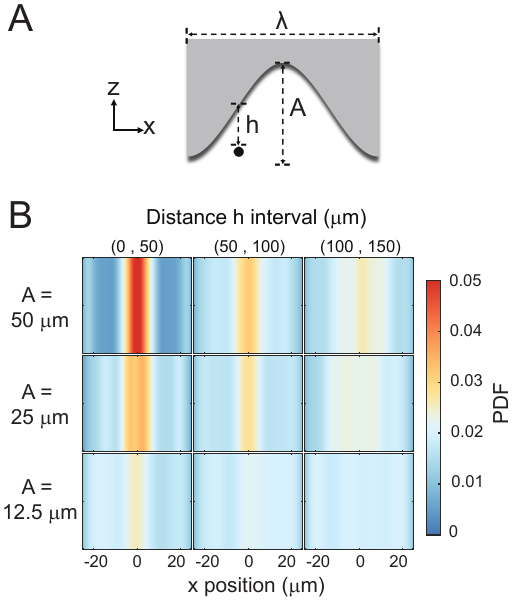}
\caption{\label{fig2}Residence time distribution of cells along the $x$-axis at different microstructure amplitudes ($A = 50~\mu\mathrm{m}$, $25~\mu\mathrm{m}$, $12.5~\mu\mathrm{m}$) with a wavelength $\lambda = 50~\mu\mathrm{m}$. Distributions are shown for different distances $h$: $0~\mu\mathrm{m} \leq h \leq 50~\mu\mathrm{m}$, $50~\mu\mathrm{m} \leq h \leq 100~\mu\mathrm{m}$, and $100~\mu\mathrm{m} \leq h \leq 150~\mu\mathrm{m}$.}
\end{figure}

{We next examine} the distribution of cell residence time in different regions near the microstructure. By applying a periodic transformation to the cell position along the $x$-axis, defined as $z(x+\lambda)=z(x)$, all cell trajectories are effectively mapped into one single period along the $x$-axis. Experimental results show that cells tend to accumulate at the valley of the microstructure rather than at the peak. To quantify this behavior, we analyze the distribution of cell residence times along the $x$-axis within intervals at different distances $h$ from the microstructure shown in Fig. \ref{fig2}A. Here, the residence time is defined as the total duration a cell remains within a specified spatial interval during its trajectory. The distance between the cells and the microstructure, defined as $h = \left| z_{\text{cell}} - z(x) \right|$, 
can be determined from the three-dimensional trajectories of cellular locomotion acquired with a three-dimensional tracking microscope.
To characterize the influence of microstructure depth on residence time, we examine three microstructures with distinct depths ($A$), tracking approximately 120 individual cells under each experimental condition, with each cell trajectory recorded for 20 seconds. As shown in Fig. \ref{fig2}B, when $0 < h < 50$, cells spend a significant amount of time near the valley of the microstructure, with the longest residence time observed for the microstructure with a depth of 50~$\mu$m. As $h$ increases, the relative residence time at the valley gradually decreases. For distance $h$ greater than 200~$\mu m$, the spatial variation in residence time across the microstructure becomes negligible, indicating a nearly uniform distribution. Crucially, our position-resolved analysis shows stronger accumulation and guidance at valleys than at peaks, refining earlier global comparisons between concave and flat features~\cite{PhysRevE.99.052607}.

\begin{figure*}
\centering
\includegraphics[width=1.0\linewidth]{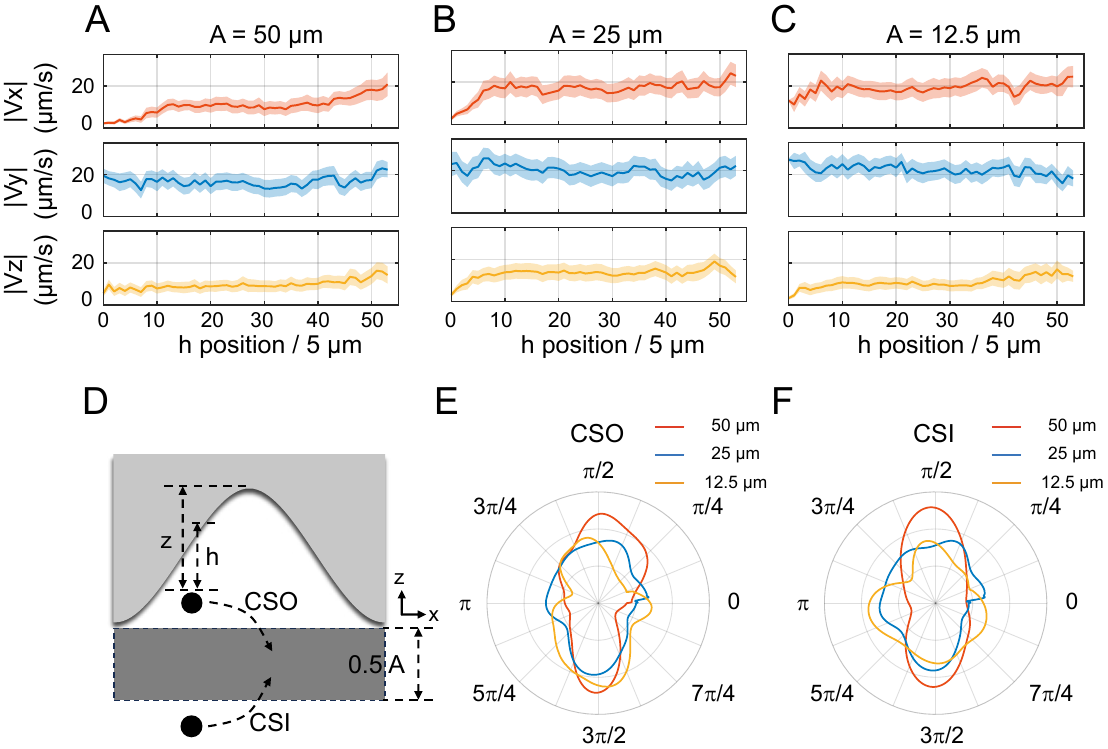}
\caption{\label{fig3}(A-C) The relationship between the magnitude of the three - dimensional spatial velocity components of cells near microstructures with depths of 12.5 $\mu m$, 25 $\mu m$, and 50 $\mu m$ and h. The x - and z - axis velocity components increase with h, while the y - axis component remains relatively stable. (D) Schematic for studying the swimming direction distribution of cells exiting and entering microstructures. For both CSO (E) and CSI (F) under different depth A, the swimming direction concentrates near $\pi/2$ and $3\pi/2$.}

\end{figure*}

\subsection{\label{sec:level2}Swimming speed near the microstructure}
To elucidate the underlying reasons of the observed residence time distribution, we examine the spatial velocity components of cells swimming near the microstructure. We find that when $0~\mu\mathrm{m} \leq h \leq 50~\mu\mathrm{m}$, velocity components along the $x$ and $z$ directions ($V_x$ and $V_z$) are relatively small compared to those along the $y$ direction ($V_y$). As shown in Fig .\ref{fig3}A-C, the magnitudes of the velocity components along the x and z directions increase with h, reaching around 20 $\mu m/s$; however, the magnitude of the velocity component along the y direction is hardly affected by h, remaining near 20 $\mu m/s$.

This difference reflects the geometric constraints and hydrodynamic interaction near the microstructure. The $x$ direction is limited by lateral confinement from the periodic surface, while the $z$ direction is influenced by both vertical confinement and wall-induced hydrodynamic interactions. In contrast, the $y$ direction remains mostly unconfined, allowing cells to sustain their swimming unaffected. To further elucidate how cells interact with the microstructure, it is necessary to distinguish between cells swimming toward the structure and those swimming away from it. These two cases may involve different swimming strategies, bounded memory effect, and hydrodynamic interaction. Therefore, in the following analysis, we separately examine the incoming and outgoing trajectories to better capture the direction-dependent behaviors and uncover the underlying mechanisms driving cell–structure interactions.

It is observed that the angle distribution between the cell swimming direction and the positive $x$-axis are not uniform, regardless of whether the cells are swimming out of the microstructure or returning to it from the outside. As shown in Fig.~\ref{fig3}D, we define the upper portion of the microstructure—approximately half of its depth—as the ``observation area''. Cells exiting the microstructure are categorized as ``Cases Swimming Outward'' (CSO), while those re-entering it are labeled as ``Cases Swimming Inward'' (CSI). In both CSO and CSI, cells display a pronounced directional bias in swimming near microstructures with all three tested depths (see Fig.~\ref{fig3}E and F). The dominant swimming orientations in both CSO and CSI cases cluster around $\pi/2$ and $3\pi/2$ relative to the $x$-axis, suggesting that the orientation of cells near the microstructure is influenced not only by random motion but also by hydrodynamic interactions. 

\begin{figure*}
\centering
\includegraphics[width=1\linewidth]{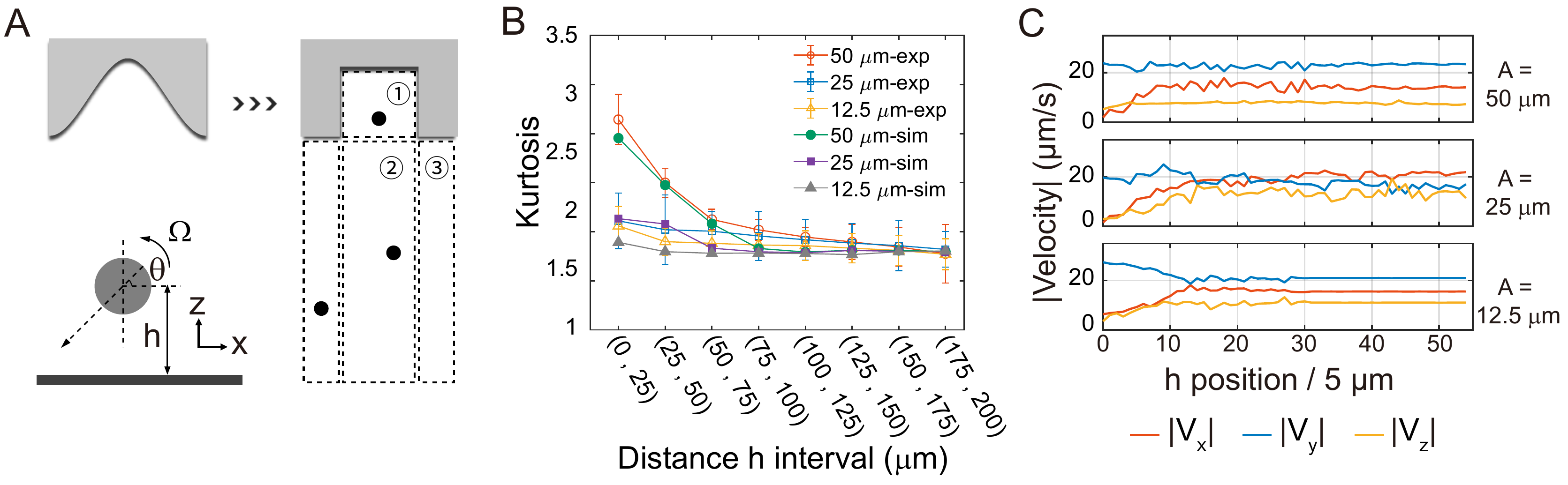}
\caption{\label{fig4}(A) Using inverse Fourier series transform to simplify the microstructure into a square - wave shape, and the schematic diagram of the hydrodynamic attraction and torque experienced by cells near the interface. (B) Kurtosis of cell residence time along the x - axis near microstructures under the combined action of contact force and hydrodynamic interaction, including comparisons between experimental and simulation results. (C) Simulation results of velocity magnitude under different typical depths.}
\end{figure*}

\subsection{\label{sec:level2}Contact and hydrodynamic interaction}
To better understand the experimental observation, we perform a simulation of the phenomenon by considering the interaction between wall-induced cell scattering and hydrodynamic effects~\cite{berke_hydrodynamic_2008,PhysRevE.98.012603}. Given the linearity of Stokes flow at low Reynolds numbers, the sinusoidal geometry of the periodic surface is simplified to a ``square-wave like" via inverse Fourier transformation (see Supplemental Material~\cite{Supplemental Material} for more details). As shown in Fig.~\ref{fig4}A, we divide the region where cells stay into three parts. In the first region, swimmers are influenced by three walls, whereas in the remaining regions, swimmers are affected by only a single wall. By incorporating the relationship between the given incidence and exit angles, along with the influence of velocity and torque exerted by the surrounding fluid during cell motion, we obtain simulation results that show good agreement with the experimental data. According to previous studies \cite{buchner_hopping_2021, kantsler_ciliary_2013,contino_microalgae_2015}, if the cell does not make  direct contact with the wall surface, the magnitude of the incidence angle is approximately equal to that of the exit angle. If the cell makes contact with the wall surface, the magnitude of the incidence angle is approximately twice that of the exit angle. Therefore, both modes are incorporated into our simulation, with $\pi/6$ taken as the threshold~\cite{buchner_hopping_2021}: when the magnitude of the incidence angle exceeds $\pi/6$, it is considered the contact mode, and the magnitude of the incidence angle is defined as twice that of the exit angle. Otherwise, it is considered the non-contact mode, in which the hydrodynamic interaction with the wall primarily governs the near-wall motion of the cell, and the magnitude of the exit angle is set equal to that of the incidence angle. 

For the influence of velocity and torque exerted by the surrounding fluid during cell motion, based on our previous work, the flow field induced by a Stokeslet dipole near a solid surface can be mathematically expressed as {\cite{berke_hydrodynamic_2008,Spagnolie2012JFM}}:  
\begin{equation}
\boldsymbol U(\boldsymbol h)=\frac {p}{8\pi\eta h^3}\left[1-3\sin^2\theta\right]\boldsymbol h
\label{U}
\end{equation}
where $p$ is the strength of the force dipole, $\theta$ and $h$ are the angle and distance between the force dipole and the surface, and $\eta$ is the fluid viscosity. Besides, the rotation effect can be expressed as
\begin{equation}
\Omega_z(\theta,h)=-\frac{3p\cos\theta\sin\theta}{64\pi\eta h^3}\left[1+\frac{(\gamma^2-1)}{2(\gamma^2+1)}(1+\sin^2\theta)\right]
\end{equation}
where $\gamma$ is the aspect ratio of the dipole. 
We simulate the near-wall locomotion of cells for approximately 40 seconds, obtain the residence time of cells at different positions along the $x$-axis, and compare the simulation results with experimental observations.

Kurtosis is used to characterize the residence time distribution of cells at different positions within the microstructure. Subsequently, we examine how kurtosis varies across three microstructures with the same period but different depths. As shown in Fig.~\ref{fig4}B, for the microstructure with a depth of 50~$\mu$m, the kurtosis of the cell residence time distribution decreases significantly with increasing distance, dropping from 3.34 to 1.8, which indicates a transition toward a uniform distribution. In contrast, for microstructures with depths of 25~$\mu$m and 12.5~$\mu$m, the change in kurtosis is less pronounced, gradually decreasing from approximately 2.1 to 1.8. These results indicate that the depth of the microstructure has a significant effect on the residence time of cells, with greater depths leading to longer cell residence times near the bottom of the structure. Based on the simulation results shown in Fig.~\ref{fig4}B, Fig.~\ref{fig4}C presents the variation of the spatial velocity components of the swimmers with height $h$, which is in good agreement with the experimental observations shown in Fig.~\ref{fig3}A-C.

\begin{figure*}
\centering
\includegraphics[width=1.0\linewidth]{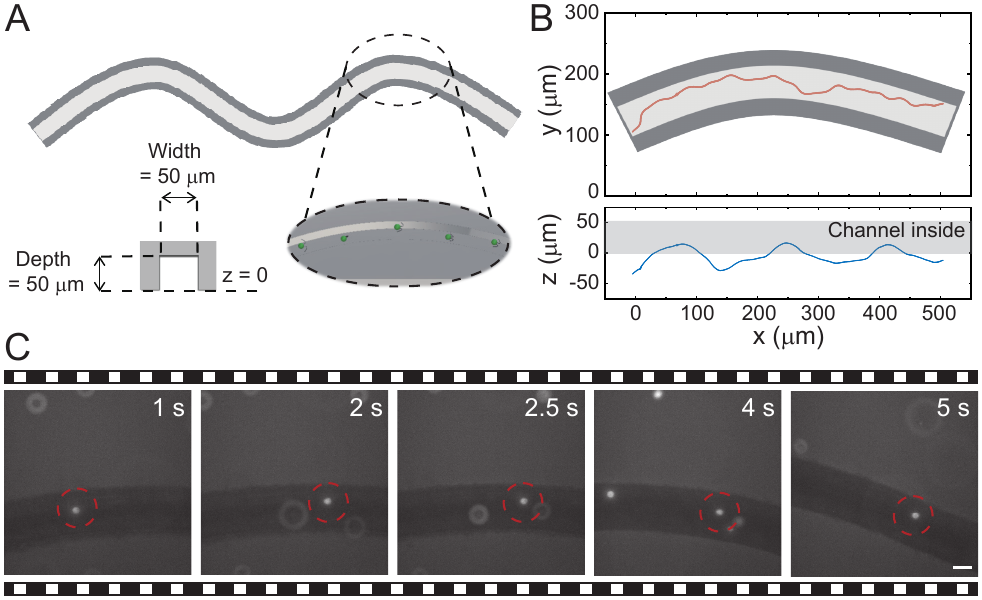}
\caption{\label{fig5}(A) Schematic of the directional guidance of cell locomotion by microstructures of specific shapes. (B) The x-y projection and x-z projection trajectory of cell locomotion , which show a ``hopping" behavior during cell locomotion. (C) Microscope snapshots of cell locomotion near microstructures at different times. The results show that the direction of cell locomotion is consistent with the shape of the microstructures. Scale bar: 20 $\mu m$. }
\end{figure*}

\subsection{\label{sec:level2}Memory effect}
We further show that directed swimming arises chiefly from microstructure-induced hydrodynamic attraction and orientation ``memory effect", not from curvature, differentiating our results from curvature-centric interpretations. The ``memory effect" refers to the persistence of swimming orientation after the cells interact with the microstructured surface. Upon encountering geometrical constraints, the cells do not immediately randomize their locomotion direction; instead, they tend to maintain a preferred orientation for a certain duration, especially when swimming along the groove direction of the microstructure. According to Fig.~\ref{fig3}A--C, it is evident that before the cell exits the microstructure, its swimming velocity component in the $y$-direction remains relatively large. Since the speed of the cell is relatively a constant, it can be inferred that when the cell just exited the microstructure, it is more likely to move along the y-direction \cite{C6SM01424K}. In addition, hydrodynamic interactions between the swimming cell and the nearby surface generate attractive forces that guides the cell’s trajectory into a hopping motion near the microstructure as illustrated in Fig.~\ref{fig5}B. Based on our previous work, we find that the interaction between $C.\ \textit{reinhardtii}$ and the surrounding liquid is not a simple ``puller" type dipole, but a composite dipole of ``puller" and ``pusher" \cite{PhysRevE.111.L052401}. The ``pusher" dipole causes the cell to be attracted to the solid-liquid interface, thus swimming towards the solid-liquid interface direction~\cite{PhysRevX.7.011010,SoftMatter.15.3397,AEM.63.3474}. These two effects jointly contribute to the anisotropic distribution of swimming orientations observed in our experiments. For cells that have just exited the microstructure, the swimming direction retains the memory from within the , still biased towards the y-axis direction. Therefore, under the influence of the ``memory effect" and hydrodynamics interaction, the cell swims along the y-axis direction while also being attracted to the solid-liquid interface and moving towards the z-axis direction. When the cell comes into contact with the wall again, it moves away from the wall and then exits the microstructure again. In this process, $V_y$ is still relatively large, and the cell is still attracted to the wall due to the ``pusher" dipole effect. In this cycle, ultimately, even when the cell is outside the microstructure, the y-axis direction speed component remains relatively large, which is consistent with the experimental results in Fig. \ref{fig1}B.

With these results, we achieve directed swimming guidance of cells by modifying the geometry of the periodic microstructure. As shown in Fig.~\ref{fig5}A, curved groove tracks with a depth and width of 50~$\mu$m are fabricated on a smooth glass surface. We then observe the swimming behavior of cells near the tracks within the chamber. Fig.~\ref{fig5}B presents the three-dimensional cell trajectory projection onto the $x$-$y$ and $y$-$z$ plane. It is clearly observed that the cell exhibits a ``hopping behavior" while swimming along the track, which is consistent with the hydrodynamic attraction model described earlier. Fig.~\ref{fig5}C shows a microscope image of a cell swimming along the track (see video in Supplemental Material~\cite{Supplemental Material}). {We find that both corrugated and square-wave-like complex surfaces can influence the swimming direction of cells. Therefore, in the future, we plan to explore how the movement trajectories and swimming behaviors of nearby cells change by adjusting the curvature and shape of the valleys on complex surfaces.}

\section{\label{sec:level1}Conclusions}
In this work, we observe the swimming direction tendency of $C.\ \textit{reinhardtii}$ near a sinusoidal periodic microstructure. {This effect persists until the distance between the cell and the microstructure exceeds twice the microstructure depth.} We also find that cells tend to stay at the valley of the  rather than at the peak of the microstructure. By resolving valley–peak differences we provide a local, position-dependent picture of concavity, complementing earlier global metrics. Regarding the directional swimming of cells, we attribute it to the combined action of orientation ``memory'' upon leaving the microstructure and hydrodynamic attraction near the solid--liquid interface. We explicitly examine this near-wall hydrodynamics as part of the mechanism, thereby extending purely steric/Brownian models. As a verification of our speculation, we successfully achieve directed induction of swimming cells near arbitrary patterned surfaces. {The study of the swimming behavior of pusher swimmers, such as \textit{E. coli}, near complex surfaces has been a long-standing research focus. However, previous simulation and numerical studies have indicated that significant hydrodynamic interactions only arise when the characteristic dimensions of the microstructure are comparable to or within the same order of magnitude as the microswimmer's size. Therefore, our future goal is to use smaller characteristic-sized complex surfaces (e.g., $\lambda = 5~\mu$m) to observe the swimming behavior of \textit{pusher swimmers} near these surfaces and analyze their dynamic features. Additionally, observing the motion of microorganisms with more complex swimming patterns near complex surfaces is also part of our future plan. Due to the more intricate flagellar motions, this will lead to a more complex flow field, making this part of the study particularly challenging.} 

{We believe that this work contributes to further exploring the motion mechanisms of microorganisms near solid-liquid interfaces, and it also has certain practical applications. For example, designing micro-robots with autonomous navigation and self-driving capabilities for micro-drug delivery or precise manipulation at the cellular level; preparing specific surface microstructures on microfluidic chips to control cell movement behavior, thereby improving the performance of the microfluidic chip, etc.}

\section{\label{sec:level1}Acknowledgments}
We thank Hepeng Zhang and Shuo Guo for the comments and discussions on the paper content. {This work is funded by NSFC 12202275, STCSM 22YF1419800.}

{
\section{Materials and Methods}}
{
\subsection{Cell Culture}}
{
$C.\ \textit{reinhardtii}$ cells (GY-D55) used in the experiments were obtained from Shanghai Guangyu Biological Technology Co., Ltd. The species were cultured in TAP (Tris-Acetate-Phosphate) Broth during the exponential growth period for 10 d to reach the cell density of 1e6 cells/mL counted by a hemocytometer (Watson 177-112C) and then exposed to the 12h-12h light/dark cycle. The temperature was 25 ℃ no matter the light cycle or the dark cycle with the illumination between 2000 and 6000 lux using white LEDs. }
{
\subsection{Chamber preparation}}
{
Circular chambers made of Polydimethylsiloxane (PDMS) were manufactured by mixing and thoroughly stirring the base and curing agent in a 10:1 mass ratio. After mixing, the mixture was placed in a vacuum chamber for degassing and then carefully pouring between two slides separated by seven spacers, each of which was 100 $\mu m$ thick. All parts were heated in an oven at 80 °C for 2 hours to complete the curing of PDMS. Removing the PDMS film from the oven and cutting neatly for subsequent experiment. At last, the PDMS film was perforated using a 1mm diameter punch.}
{
\subsection{Fluorescence Tracking Microscopy}}
{
Real-time 3D tracking technology is employed to continuously record three-dimensional swimming trajectories of individual $C.\ \textit{reinhardtii}$ for extended periods (approximately 30 seconds). An inverted microscope (Nikon ECLIPSE Ti2-U) equipped with a PX.EDGE 5.5-sCMOS camera captures cell images. The tracking experiments take place in a dark chamber. A 565~nm wavelength LED light source is used to excite \textit{C. reinhardtii}, and the cells emit fluorescence around 610~nm. A two-dimensional displacement stage tracks cellular locomotion along the plane of parallel glass plates, while a piezoelectric displacement stage (Pi) simultaneously records vertical motion and controls focal plane positioning to maintain continuous cell visibility. A custom C++ algorithm acquires images at 60 fps (300×300 pixel resolution) while synchronously recording data from both displacement stages. This system ultimately generates real-time three-dimensional motion trajectories of cells within the specified timeframe~\cite{qu_changes_2018}. }
{
\subsection{Inverse Fourier Transform}}
{
It is well known that any periodic function can be represented as a Fourier series consisting of sine and cosine terms. In this study, we consider a square-wave function as the target function, with wavelength $\lambda = 50~\mu\mathrm{m}$. Triangular functions with amplitudes $A = 50~\mu\mathrm{m}$, $25~\mu\mathrm{m}$, and $12.5~\mu\mathrm{m}$ are used as the corresponding first-order approximations of the square-wave function. This allows us to employ the mirror-image method \cite{berke_hydrodynamic_2008,PhysRevE.111.L052401} to model the attractive forces and torques experienced by cells near the wall.}
{
Specifically, the triangular function
\begin{equation}
f(x) = A \sin\left(\frac{\pi}{25} x - \frac{\pi}{2}\right)
\end{equation}
can be approximated by the square-wave function
\begin{equation}
f(x) =
\begin{cases}
A, & x \in [-25, -12.5) \cup (12.5, 25] \\
-A, & x \in [-12.5, 12.5]
\end{cases}.
\end{equation} 
}

\nocite{*}


\end{document}


\maketitle

\begin{figure}[!htbp]
    \centering
    \includegraphics[width=0.8\linewidth]{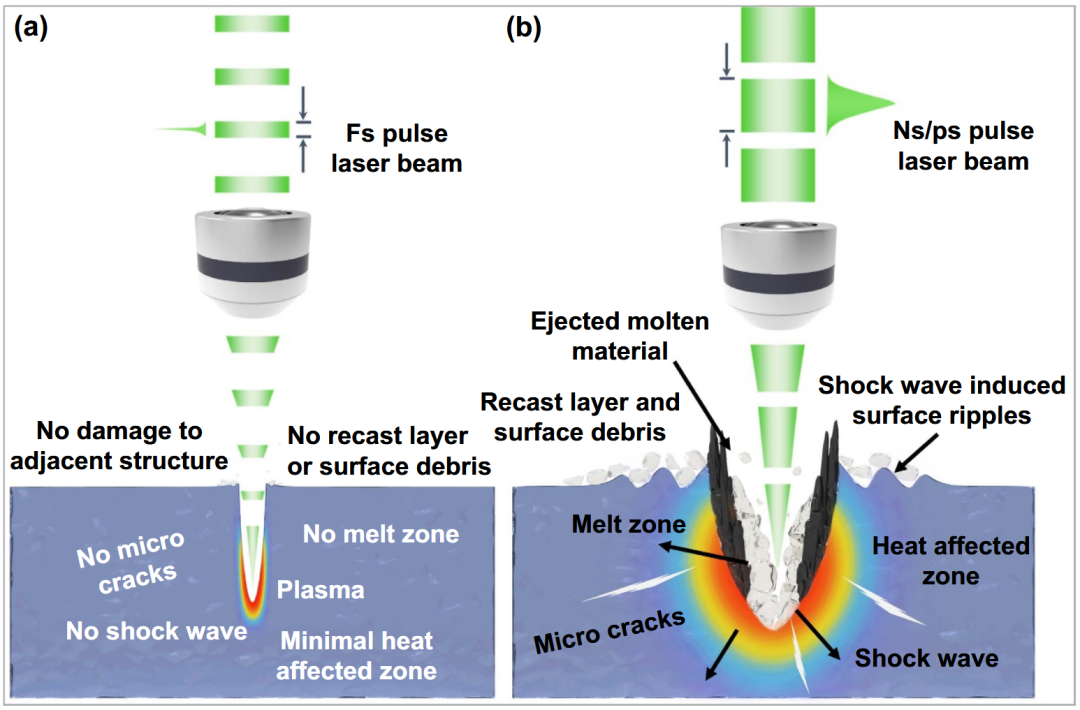}\\
	\caption{Comparison between femtosecond and nanosecond/picosecond laser processing. 
(a) Femtosecond laser pulses induce material modification with negligible surface damage and an extremely small heat-affected zone. 
(b) In contrast, nanosecond/picosecond laser irradiation typically leads to pronounced thermal damage and defect formation in the processed material~\cite{Gao2025FemtosecondLaser}.}
	\label{complex_surface}
\end{figure}

\begingroup

\section{\label{sec:level1}Complex microstructure preparation}
In this work, the corrugated surfaces used in the experiments are fabricated by Laserrush (Nanjing, China) using femtosecond laser processing. This technique uses extremely short laser pulses to shape materials with very high precision. A key feature of femtosecond laser processing is that the laser pulse is so brief that the material does not have time to heat up significantly during the interaction. As a result, the material is modified before heat can spread to the surrounding region. This behavior is often referred to as “cold processing” (Fig.~\ref{complex_surface}). Because heating is strongly suppressed, the affected area remains highly localized and the surrounding material is largely undamaged. This stands in contrast to conventional continuous-wave or nanosecond lasers, which typically generate substantial heat and can cause melting, resolidification, or microcracks. As a consequence, femtosecond laser processing allows precise control over surface geometry while preserving the structural integrity of the material. This makes it particularly suitable for fabricating well-defined micro- and nano-scale surface structures, which is essential for the experiments presented in this thesis.

\section{\label{sec:level1}Transformation from the $z$-coordinate to the distance $h$ relative to the microstructure}
We model the microstructure profile using the function
\begin{equation}
h' = A \sin\left(\frac{2\pi}{\lambda}x + \varphi\right) + C,
\end{equation}
as illustrated in Figure S1A. For each 3D fluorescence trajectory, we set the initial phase to \(\varphi = -\pi/2\) at the first frame. For subsequent frames, the two bright boundaries of the microstructure allow determination of \(a_1\), \(a_2\), and the cell's x-coordinate \(b\) (Figure S1B). It follows that \(|a_1 - a_2| = \lambda\).

Using image analysis for the \(n\)-th frame, we can express the height as
\begin{equation}
h' = A \sin\left[\frac{2\pi (b - a_1)}{a_2 - a_1}\right] + C = A \sin\left(\frac{2\pi x}{\lambda} + \varphi \right) + C,
\end{equation}
from which the phase \(\varphi\) can be obtained:
\begin{equation}
\varphi = \frac{2\pi (b - a_1)}{a_2 - a_1} - \frac{2\pi x}{\lambda}.
\end{equation}
This allows calculation of \(h'\) for each frame. Finally, the distance from the cell to the microstructure surface is computed as
\begin{equation}
h = H - z - h',
\end{equation}
where \(H\) is the chamber thickness and \(z\) is the original z-coordinate.

\section{\label{sec:level1}Analysis of the skewness of $V_x$ within the target area}
We also identified experimental evidence supporting the tendency of cells to accumulate in the valleys of the microstructure. Firstly, we apply a periodic transformation, \( z(x+\lambda) = z(x) \), combined with a symmetry transformation, to map all cell trajectories into a single half-period of the microstructure. We then define a ``Target area'' as the central region with a width of \(0.25\lambda\) (measured along the \(x\)-direction) and analyze the distribution of \(V_x\) for cells located within this region. The skewness \(S\) of the \(V_x\) distribution is calculated as  
\begin{equation}
S = \frac{\frac{1}{N} \sum_{i=1}^N \left(V_{x,i} - \overline{V_x}\right)^3}{\left[ \frac{1}{N} \sum_{i=1}^N \left(V_{x,i} - \overline{V_x}\right)^2 \right]^{3/2}},
\end{equation}
where \(N\) is the number of data points and \(\overline{V_x}\) is the mean \(V_x\) value. A negative skewness indicates that the distribution has a longer tail on the left side (toward smaller \(V_x\) values), while a positive skewness indicates a longer tail on the right side (toward larger \(V_x\) values).

The results show that positive \(V_x\) values occur more frequently, and the corresponding \(V_x\) distributions exhibit negative skewness, indicating a statistical bias toward rightward motion. We compute the skewness of the \(V_x\) distribution in the Target area for cells swimming near microstructures of varying depths at different heights \(h\). For \(h < 100~\mu\text{m}\), the distributions are negatively skewed, implying a preferential movement toward the right, which is consistent with the experimentally observed accumulation of cells in the valleys of the microstructure. For microstructures with depths of \(25~\mu\text{m}\) and \(12.5~\mu\text{m}\), this negative skewness disappears when \(h > 100~\mu\text{m}\). In contrast, for microstructures with a depth of \(50~\mu\text{m}\), the disappearance of negative skewness occurs only when \(h > 150~\mu\text{m}\).

 \endgroup

\clearpage

\begin{figure}[!htbp]
    \centering
    \includegraphics[width=0.8\linewidth]{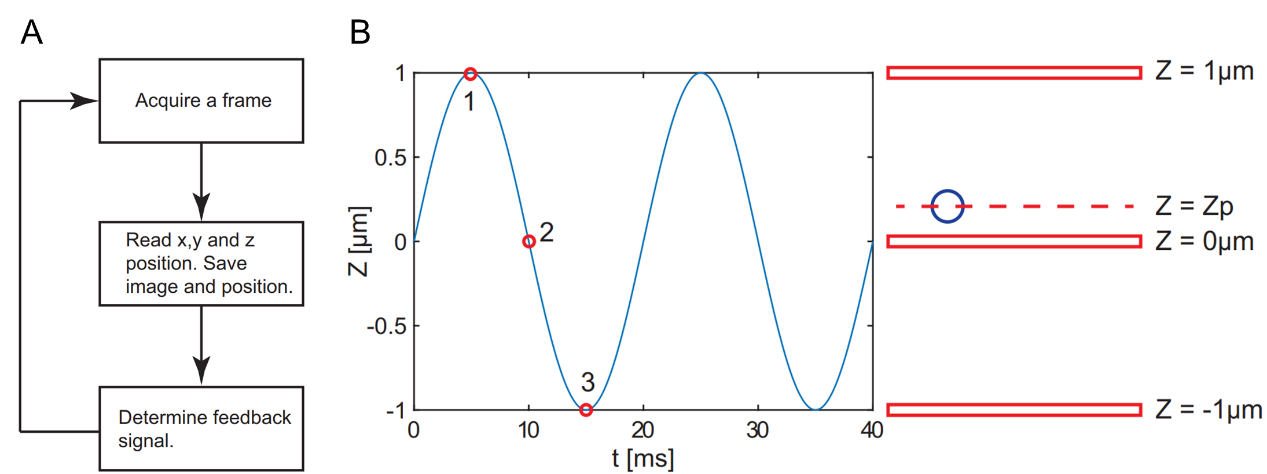}\\
	\caption{(A) Flow chart of real-time three-dimensional tracking algorithm. (B) A demonstration of tracking along optical axis using active scanning. Three consecutive images are taken at location 1, 2 and 3. Intensity measurements from those three images lead
to an accurate detection of particle location in z~\cite{Thesis.2018.Qu}.}
	\label{piezo}
\end{figure}

\begin{figure}[htbp]
\centering
\includegraphics[width=\textwidth]{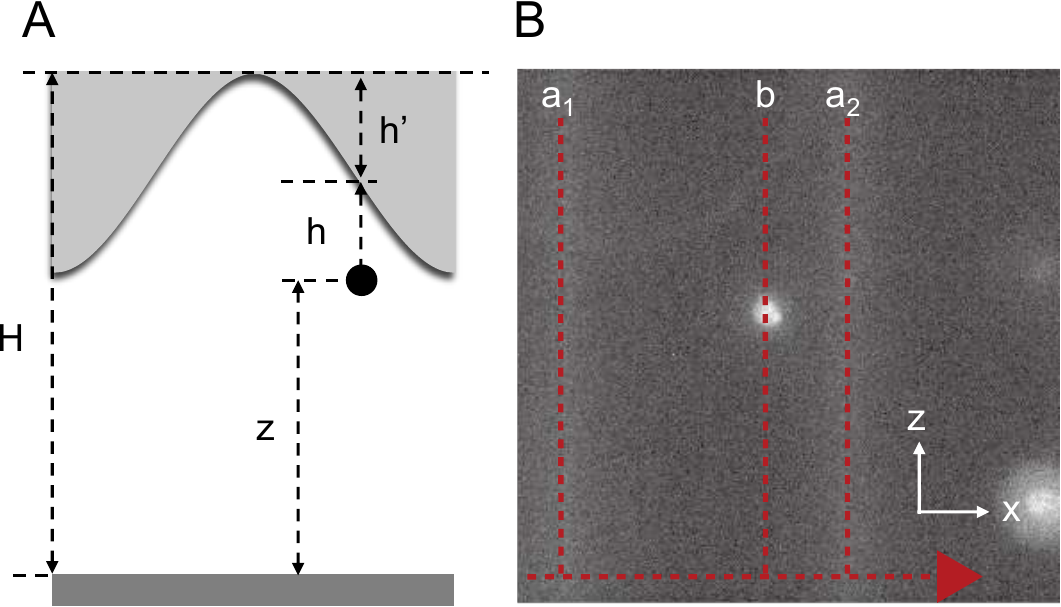}
\caption{(A) Schematic illustration of the derivation of the cell-to-microstructure distance, $h$.  
(B) A representative fluorescence image showing the positions corresponding to $a_1$, $a_2$, and $b$.}
\end{figure}

\begin{figure}[htbp]
\centering
\includegraphics[width=\textwidth]{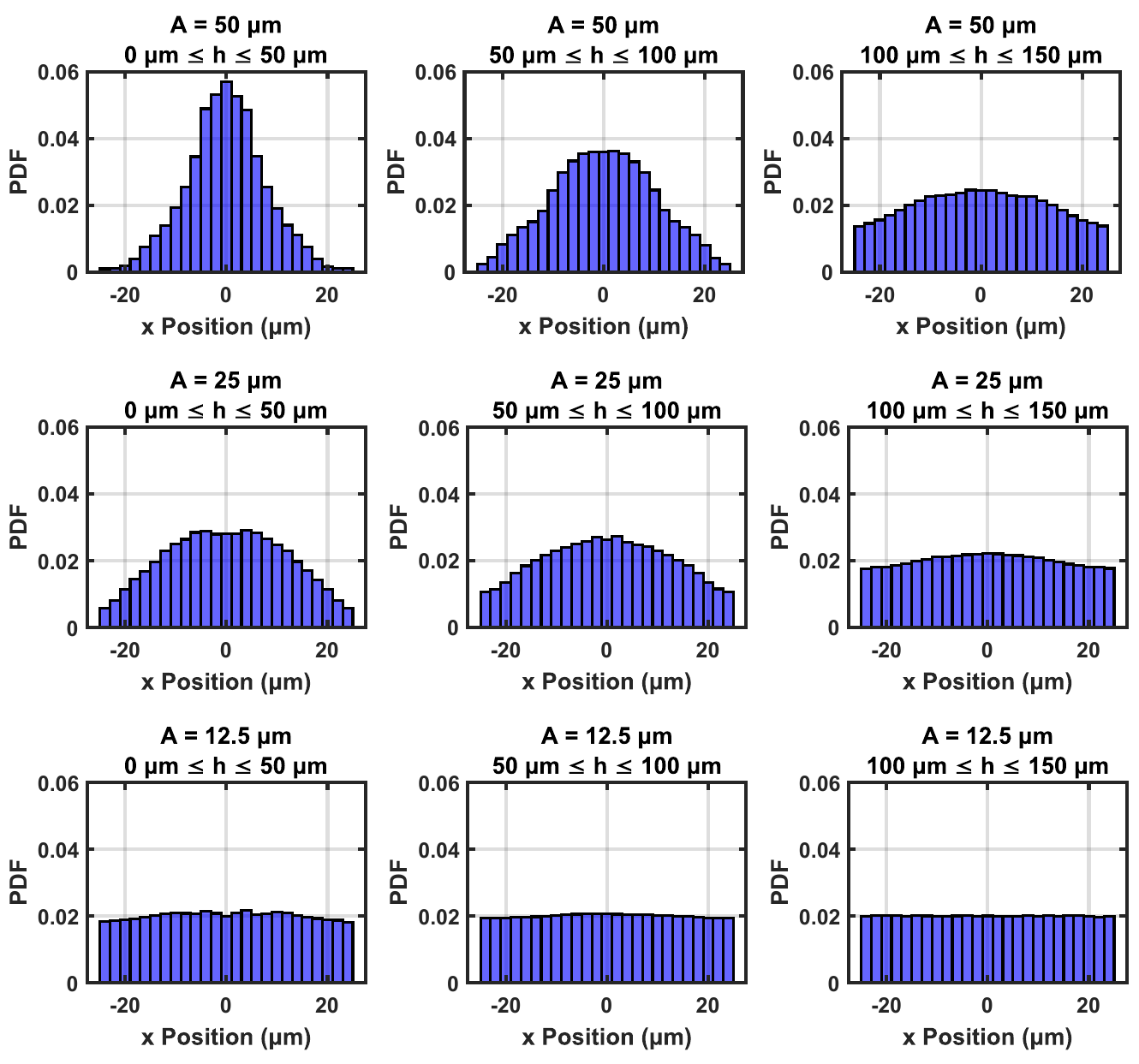}
\caption{\label{sim_residence_time}Simulation result of residence time distribution of cells along the $x$-axis at different microstructure amplitudes ($A = 50~\mu\mathrm{m}$, $25~\mu\mathrm{m}$, $12.5~\mu\mathrm{m}$) with a wavelength $\lambda = 50~\mu\mathrm{m}$. Distributions are shown for different distances $h$: $0~\mu\mathrm{m} \leq h \leq 50~\mu\mathrm{m}$, $50~\mu\mathrm{m} \leq h \leq 100~\mu\mathrm{m}$, and $100~\mu\mathrm{m} \leq h \leq 150~\mu\mathrm{m}$.}
\end{figure}

\begin{figure}[htbp]
\centering
\includegraphics[width=\textwidth]{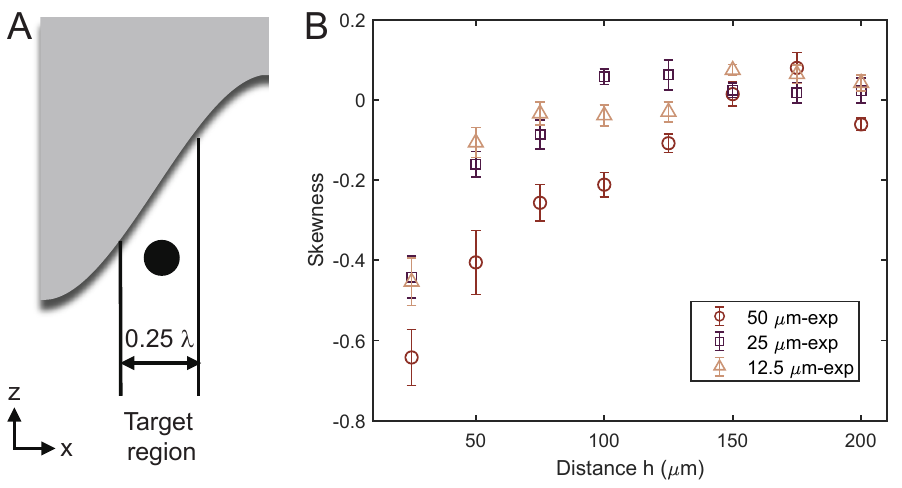}
\caption{Analysis of cell swimming near microstructures. 
(A) Schematic illustration of a single half-period of the microstructure and the defined ``Target area'' (central region with width $0.25\lambda$) used for analysis of $V_x$ distributions. 
(B) Skewness of $V_x$ within the Target area as a function of height $h$ for microstructures of different depths. Negative skewness indicates a left-tailed distribution corresponding to preferential rightward cell motion.}
\end{figure}

\begin{figure}[htbp]
\centering
\includegraphics[width=\textwidth]{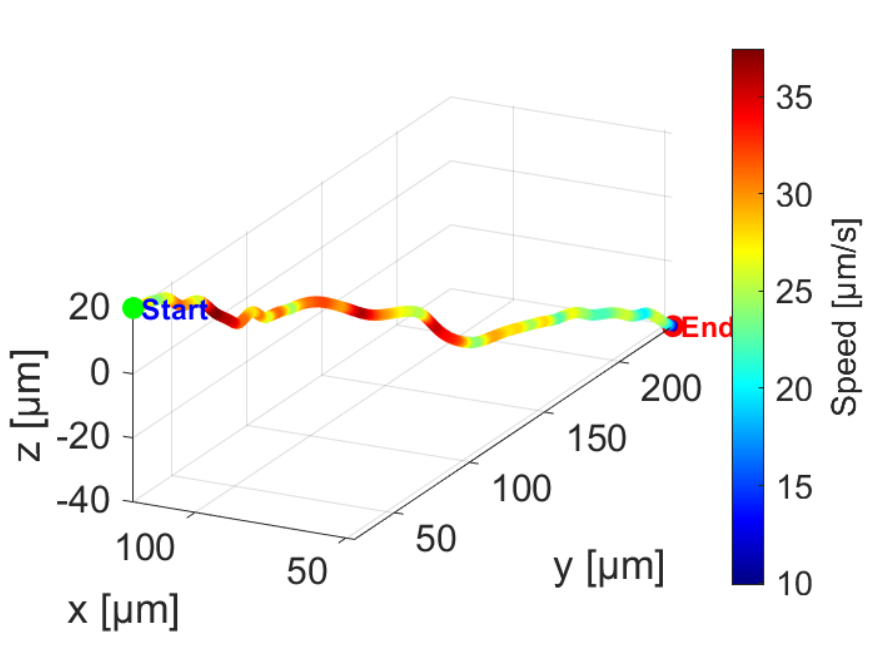}
\caption{Three-dimensional trajectory of $C. \textit{reinhardtii}$ using the 3D real-time fluorescence tracking microscope, and colorbar represents the cell swimming speed.}
\end{figure}

\clearpage 

\clearpage
\paragraph{Caption for Movie S1.}
\textbf{Directed swimming behavior of one typical cell along an S-shaped track}

\paragraph{Caption for Movie S2.}
\textbf{Directed swimming behavior of one typical cell along a zigzag track}